\documentclass[12pt]{article}
\usepackage{amssymb, amsmath, bm, float, epsfig, color}

\begin{document}
\title{
\vspace{-20pt}
\begin{flushright}
\normalsize WU-HEP-16-14 \\
HUPD1606 \\*[55pt]
\end{flushright}
{\Large \bf 
Edge of a cliff
\\*[20pt]}
}

\author{
Yusuke~Shimizu,$^1$\footnote{E-mail address: yu-shimizu@hiroshima-u.ac.jp} \, and \,
Yoshiyuki~Tatsuta$^2$\footnote{E-mail address: y\_tatsuta@akane.waseda.jp}\\*[30pt]
$^1${\it \normalsize Graduate~School~of~Science,~Hiroshima~University,~Higashi-Hiroshima,~739-8526,~Japan}\\
$^2${\it \normalsize Department~of~Physics,~Waseda~University,~Tokyo~169-8555,~Japan}\\*[55pt]
}

\date{
\centerline{\small \bf Abstract}
\begin{minipage}{0.9\textwidth}
\medskip\medskip 
\small
We discuss the neutrino flavor structures in the Occam's razor approach for the Dirac neutrino mass matrices. 
We assume that the charged lepton mass matrix takes a diagonal base, while the right-handed Majorana neutrino mass matrix is also diagonal and we consider nine patterns of the four zero textures for the Dirac neutrinos mass matrices. We numerically analyze the left-handed Majorana neutrino mass matrices for nine patterns of the Dirac neutrino mass matrices 
and we find two interesting patterns where both normal and inverted neutrino mass hierarchies can be realized. 
We also find if the neutrino mass is normal hierarchy, this scenario will be likely to be excluded by the measurements of the Dirac CP violating phase in the T2K and NO$\nu $A neutrino experiments, for instance. 
On the other hand, if the neutrino mass is inverted hierarchy, this scenario will be also likely to be excluded by the measurements of neutrinoless double beta decay experiments, e.g., KamLAND-Zen experiment. 
\end{minipage}}

\begin{titlepage}
\maketitle
\thispagestyle{empty}
\end{titlepage}

\section{Introduction}
Now the standard model (SM) is one of the successful models in the particle physics, completed by the recent discovery of the Higgs boson.
Almost all input parameters in the SM of the particle physics are known to be coupling constants of Yukawa-type interactions.
Such coupling constants, in other words, magnitudes of interactions among quarks/leptons and the Higgs boson that acquire the vacuum expectation value after the electroweak symmetry breaking, determine the masses and mixing angles of quarks and leptons.
Although Yukawa coupling constants are quite important from such a point of view, concrete values of Yukawa coupling constants remain still experimentally unmeasured and ambiguous.
This is called the flavor puzzle.
Recently, Tanimoto and Yanagida applied a Scholastic idea of William of Ockham, {\em ``Entities should not be multiplied unnecessarily."}, i.e., the Occam's razor, to mass matrices in the SM~\cite{Tanimoto:2016rqy}. 
The Occam's razor approach is proposed and investigated originally in the quark sector, 
where Tanimoto and Yanagida assumed the scheme for a minimum set of free parameters in the quark mass matrices in order to realize the successful quark masses, Cabbibo-Kobayashi-Maskawa (CKM) mixing, and CP violation, simultaneously. 
This Occam's razor approach is extended to the lepton sector in anticipation of leptogenesis 
in Ref.~\cite{Kaneta:2016gbq}.\footnote{The Occam's razor approach in right-handed Majorana neutrinos is also studied in Ref.~\cite{Harigaya:2012bw}.}
Then, this approach should be verified in the near future experiments. 

Neutrino oscillation experiments have provided us the important information that the two neutrino mass squared differences and the non-zero reactor angle as well as the two large lepton mixing angles. 
The Dirac CP violating phase for the lepton sector is on the next stage of the measurements by T2K and NO$\nu $A neutrino experiments \cite{Abe:2013hdq, Adamson:2016tbq}, for instance. 
If the neutrinos are Majorana particles, the Majorana phases are also sources of the CP violation. 
The KamLAND-Zen experiment \cite{KamLAND-Zen:2016pfg} which is one of the neutrinoless double beta ($0\nu \beta \beta $) decay experiments is searching for the evidence of the Majorana particles. 
It is an important task to study a new physics beyond the SM as the origin of the lepton flavor structure as well as the quark one from the theoretical point of view. 

In this paper, we discuss the neutrino flavor structures in the Occam's razor approach for the Dirac neutrino mass matrices. 
We assume that the charged lepton mass matrix takes a diagonal base, while the right-handed Majorana neutrino mass matrix is also diagonal and we consider nine patterns of the four zero textures for the Dirac neutrinos mass matrices~\cite{Branco:2007nb,Choubey:2008tb}, as the same choice as the form of the mass matrices in Ref.~\cite{Tanimoto:2016rqy}. We numerically analyze the left-handed Majorana neutrino mass matrices for nine patterns of the Dirac neutrino mass matrices 
and we find two interesting patterns where both normal and inverted neutrino mass hierarchies can be realized. 
We also find if the neutrino mass is normal hierarchy (NH), this scenario will be likely to be excluded by the measurements of the Dirac CP violating phase in the T2K and NO$\nu $A neutrino experiments, for instance. 
On the other hand, if the neutrino mass is inverted hierarchy (IH), this scenario will be also likely to be excluded by the measurements of $0\nu \beta \beta $ decay experiments, e.g., KamLAND-Zen experiment. 
Thus, these textures via the Occam's razor approach is now on the edge of a cliff.

This paper is organized as follows. In section~\ref{sec:matrix}, we discuss the neutrino mass matrices in the Occam's razor approach. 
In section~\ref{sec:analyses}, we show the numerical results for the left-handed Majorana neutrino phenomenology. 
The section~\ref{sec:summary} is devoted to discussions and summary.

\section{Neutrino mass matrix in the Occam's razor}
\label{sec:matrix}
In this section, we discuss the neutrino mass matrices in the Occam's razor approach. 
Let us discuss the lepton sector. We assume that the charged lepton mass matrix takes a diagonal base, while we also assume that the mass matrix for the right-handed Majorana neutrino $M_R$ is diagonal as
\begin{equation}
\label{eq:Majorana}
M_R=m_0
\begin{pmatrix}
\frac{1}{k_1}e^{i\phi _A} & 0 & 0 \\
0 & \frac{1}{k_2}e^{i\phi _B} & 0 \\
0 & 0 & 1
\end{pmatrix},
\end{equation}
where $m_0$ is an overall mass parameter for the right-handed Majorana neutrinos, $k_1$ and $k_2$ denote ratios among the hierarchies of the right-handed Majorana neutrinos, 
and $\phi _A$ and $\phi _B$ are complex phases which are origins of the CP violating phase~\cite{Kaneta:2016gbq}. 
Hereafter, we adopt non-vanishing $(3,3)$, $(2,3)$, and $(1,2)$ elements in the Dirac mass matrix and $\det M_D \neq 0$ in order to reproduce all non-zero neutrino masses, as adopted in Ref.~\cite{Tanimoto:2016rqy}. 
Tanimoto and Yanagida {considered} non-vanishing $(3,3)$, $(2,3)$, and $(1,2)$ elements to obtain two non-vanishing mixing angles $\theta_{12}$ and $\theta_{23}$ in the quark sector \cite{Tanimoto:2016rqy}.
We respect their {criterion} of non-zero input entries and also adopt the same non-zero entries in the neutrino Dirac mass matrix. Indeed, it is easily found that the entries are suitable also for the lepton sector.
Thanks to non-vanishing entries in $(3,3)$, $(2,3)$, and $(1,2)$ elements, a non-zero 2-3 mixing is inevitably induced in left-handed Majorana neutrino mass matrices at the low energy. 
The criterion of such three non-vanishing entries is considered to be suitable for realizing the observed large 2-3 mixing. For the criterion, the Dirac neutrino mass matrices $M_D^{(i)}~(i=1-9)$ are categorized as
\begin{align}
\label{eq:Dirac}
M_D^{(1)}&=
\begin{pmatrix}
b' & a & 0 \\
a' & 0 & b \\
0 & 0 & c
\end{pmatrix},\quad 
M_D^{(2)}=
\begin{pmatrix}
0 & a & b' \\
a' & 0 & b \\
0 & 0 & c
\end{pmatrix},\quad 
M_D^{(3)}=
\begin{pmatrix}
0 & a & 0 \\
a' & b' & b \\
0 & 0 & c
\end{pmatrix}, \nonumber \\
M_D^{(4)}&=
\begin{pmatrix}
0 & a & 0 \\
a' & 0 & b \\
b' & 0 & c
\end{pmatrix},\quad 
M_D^{(5)}=
\begin{pmatrix}
0 & a & 0 \\
a' & 0 & b \\
0 & b' & c
\end{pmatrix},\quad 
M_D^{(6)}=
\begin{pmatrix}
b' & a & 0 \\
0 & 0 & b \\
a' & 0 & c
\end{pmatrix}, \nonumber \\
M_D^{(7)}&=
\begin{pmatrix}
0 & a & b' \\
0 & 0 & b \\
a' & 0 & c
\end{pmatrix},\quad 
M_D^{(8)}=
\begin{pmatrix}
0 & a & 0 \\
0 & b' & b \\
a' & 0 & c
\end{pmatrix},\quad 
M_D^{(9)}=
\begin{pmatrix}
0 & a & 0 \\
0 & 0 & b \\
a' & b' & c
\end{pmatrix},
\end{align}
where $a$, $a'$, $b$, $b'$, and $c$ are mass parameters and can be taken to be real for field redefinitions without loss of generality~\cite{Kaneta:2016gbq}.
Here we define the $LR$ base in the Dirac mass matrices.
Note that $M_D^{(4)}$ case is excluded by current experimental data~\cite{Gonzalez-Garcia:2015qrr} because only 2-3 mixing is obtained and 
$M_D^{(5)}$ case has already studied in Ref.~\cite{Kaneta:2016gbq}.
By using seesaw mechanism~\cite{Minkowski:1977sc}-\cite{Schechter:1981cv}, we obtain left-handed Majorana neutrino mass matrices $M_\nu ^{(i)}\simeq M_D^{(i)}M_R^{-1}M_D^{(i)T}~(i=1-9)$, 
e.g., $M_D^{(2)}$ and $M^{(7)}_D$ case in Eq.~(\ref{eq:Dirac}) as follows:
\begin{align}
M_\nu ^{(2)}&=\frac{1}{m_0}
\begin{pmatrix}
a^2k_2e^{-i\phi _B}+b'^2 & bb' & b'c \\
bb' & a'^2k_1e^{-i\phi _A}+b^2 & bc \\
b'c & bc & c^2
\end{pmatrix}, \nonumber \\
M_\nu ^{(7)}&=\frac{1}{m_0}
\begin{pmatrix}
a^2k_2e^{-i\phi _B}+b'^2 & bb' & b'c \\
bb' & b^2 & bc \\
b'c & bc & a'^2k_1e^{-i\phi _A}+c^2
\end{pmatrix},
\end{align}
and we numerically analyze nine patterns of Dirac neutrino mass matrices in the next section.\footnote{There is a possibility in connecting each of the nine patterns or the other four zero textures via some unitary transformation. 
Then, it is convenient and interesting to focus only on the texture analyses of the low energy left-handed Majorana {neutrino} mass matrices. Indeed, such analyses were investigated by Merle and Rodejohann \cite{Merle:2006du}.}

Before closing this section, we list the lepton mixing matrix, the Jarlskog invariant $J_{CP}$~\cite{Jarlskog:1985ht} which is the parameter describing the size of the CP violation, 
and the effective mass of the $0\nu \beta \beta $ decay.
The lepton mixing matrix, called Pontecorvo-Maki-Nakagawa-Sakata (PMNS) matrix $U_{\text{PMNS}}$~\cite{Maki:1962mu,Pontecorvo:1967fh}, is written as 
\begin{equation}
U_{\text{PMNS}}\equiv 
\begin{pmatrix}
c_{12}c_{13} & s_{12}c_{13} & s_{13}e^{-i\delta _{CP}} \\
-s_{12}c_{23}-c_{12}s_{23}s_{13}e^{i\delta _{CP}} & c_{12}c_{23}-s_{12}s_{23}s_{13}e^{i\delta _{CP}} & s_{23}c_{13} \\
s_{12}s_{23}-c_{12}c_{23}s_{13}e^{i\delta _{CP}} & -c_{12}s_{23}-s_{12}c_{23}s_{13}e^{i\delta _{CP}} & c_{23}c_{13}
\end{pmatrix}
\begin{pmatrix}
e^{i\alpha } & 0 & 0 \\
0 & e^{i\beta } & 0 \\
0 & 0 & 1
\end{pmatrix},
\end{equation}
where $c_{ij}$ and $s_{ij}$ denote $\cos \theta _{ij}$ and $\sin \theta _{ij}$, $\delta _{CP}$ denotes the Dirac CP violating phase, and $\alpha $ and $\beta $ are Majorana phases. 
The Jarlskog invariant $J_{CP}$~\cite{Jarlskog:1985ht} is defined as 
\begin{equation}
J_{CP}=\text{Im}\left [U_{e1}U_{\mu 2}U_{e2}^*U_{\mu 1}^*\right ],
\end{equation}
where $U_{ij}$s are the PMNS matrix elements. Then, the $J_{CP}$ is explicitly written down by the lepton mixing angles and the Dirac CP violating phase as 
\begin{equation}
J_{CP}=s_{23}c_{23}s_{12}c_{12}s_{13}c_{13}^2\sin \delta _{CP}.
\end{equation}
The effective mass of the $0\nu \beta \beta $ decay is given in terms of the PMNS matrix elements and the left-handed Majorana neutrino masses $m_i~(i=1-3)$ as 
\begin{equation}
m_{ee}=\sum _{i=1}^3m_iU_{ei}^2~.
\end{equation}

\section{Numerical analyses}
\label{sec:analyses}
In this section, we analyze the left-handed Majorana neutrino mass matrices for nine patterns of Dirac neutrino mass matrices in Eq.~(\ref{eq:Dirac}).
As we have mentioned in the previous section, $M_D^{(4)}$ case is excluded by the current experimental data~\cite{Gonzalez-Garcia:2015qrr} 
and $M_D^{(5)}$ case has already studied in Ref.~\cite{Kaneta:2016gbq}.\footnote{In Ref.~\cite{Kaneta:2016gbq}, the authors had studied both of normal and inverted neutrino mass hierarchies 
and they did not find IH case. We have also studied inverted mass hierarchy in $M_D^{(5)}$ case and we found just a few allowed points. By the reproduction of the result in Ref.~\cite{Kaneta:2016gbq}, it is found that there is no characteristic prediction and testable property, even though $M_D^{(5)}$ can splendidly explain the observed 
{neutrino} mass squared differences and {lepton} mixing angles.} 
Then, we analyze the other patterns in Eq.~(\ref{eq:Dirac}). 
In our numerical calculations, $3\sigma$ ranges of mass squared differences and three mixing angles \cite{Gonzalez-Garcia:2015qrr} 
are used to cut undesired results for NH case, 
\begin{gather}
7.02 \leq \frac{\Delta m^2_{21}}{10^{-5} \, {\rm eV}^2} \leq 8.09, \qquad 2.317 \leq \frac{\Delta m^2_{31}}{10^{-5} \, {\rm eV}^2} \leq 2.607, \nonumber \\
0.270 \leq \sin^2 \theta_{12} \leq 0.344, \qquad 0.382 \leq \sin^2 \theta_{23} \leq 0.643, \qquad 0.0186 \leq \sin^2\theta_{13} \leq 0.0250,
\label{dataNH}
\end{gather}
and for IH case, 
\begin{gather}
7.02 \leq \frac{\Delta m^2_{21}}{10^{-5} \, {\rm eV}^2} \leq 8.09, \qquad -2.590 \leq \frac{\Delta m^2_{32}}{10^{-5} \, {\rm eV}^2} \leq -2.307, \nonumber \\
0.270 \leq \sin^2 \theta_{12} \leq 0.344, \qquad 0.389 \leq \sin^2 \theta_{23} \leq 0.644, \qquad 0.0188 \leq \sin^2\theta_{13} \leq 0.0251.
\label{dataIH}
\end{gather}
Indeed, since diagonal parts in the right-handed Majorana mass matrix contain an overall parameter, i.e., $m_0$, all we have to do is to use a cut for the ratio of mass squared differences.
This is because by using the overall parameter, we can determine a mass scale of the left-handed Majorana neutrinos consistent with experimental data in any cases.

The number of trials in scattering plots is $5 \times 10^{10}$. We find more predictive numerical results for $M_D^{(2)}$ and $M_D^{(7)}$ cases than those for the other cases $M_D^{(i)} \, (i=1,3,6,8,9)$ in Eq.~(\ref{eq:Dirac}). 
Hereafter we focus only on the $M_D^{(2)}$ and $M_D^{(7)}$ cases. 

Consequently, we find the following parameter regions that can realize the mass squared differences and three mixing angles for both NH and IH cases of $M_D^{(2)}$ and $M_D^{(7)}$, as shown in Tables~\ref{tab:NH} and \ref{tab:IH}, respectively. 
By using the parameter regions in Tables~\ref{tab:NH} and \ref{tab:IH}, we show their numerical results in the following.
Note that it is relatively difficult to obtain the allowed points within $3\sigma $ ranges for current experimental results~\cite{Gonzalez-Garcia:2015qrr} in IH cases than that in NH cases of $M_D^{(2)}$ and $M_D^{(7)}$.
\begin{table}[H]
\centering
\begin{tabular}{|cccccccccc|} \hline
& $a \, [{\rm GeV}]$ & $a' \, [{\rm GeV}]$ & $b \, [{\rm GeV}]$ & $b' \, [{\rm GeV}]$ & $c \, [{\rm GeV}]$ & $k_1$ & $k_2$ & $\phi_A \, [{\rm rad}]$ & $\phi_B \, [{\rm rad}]$ \\ \hline
$M_D^{(2)}$ & 0 -- 0.6 & 0 -- 0.6 & 0.2 -- 1.0 & 0 -- 0.4 & 0.3 -- 1.0 & 0 -- 500 & 1 -- 500 & $0$ -- $2\pi$ & $0$ -- $2\pi$ \\ \hline
$M_D^{(7)}$ & 0 -- 0.3 & 0 -- 0.8 & 0.3 -- 1.0 & 0 -- 0.4 & 0.4 -- 1.0 & 0 -- 500 & 1 -- 500 & $0$ -- $2\pi$ & $0$ -- $2\pi$ \\ \hline
\end{tabular}
\caption{Allowed parameter regions for NH cases of $M_D^{(2)}$ and $M_D^{(7)}$ in Eq.~(\ref{eq:Dirac}).}
\label{tab:NH}
\end{table}
\begin{table}[H]
\centering
\begin{tabular}{|cccccccccc|} \hline
& $a \, [{\rm GeV}]$ & $a' \, [{\rm GeV}]$ & $b \, [{\rm GeV}]$ & $b' \, [{\rm GeV}]$ & $c \, [{\rm GeV}]$ & $k_1$ & $k_2$ & $\phi_A \, [{\rm rad}]$ & $\phi_B \, [{\rm rad}]$ \\ \hline
$M_D^{(2)}$ & 0.3 -- 1 & 0.2 -- 0.8 & 0.2 -- 1 & 0.2 -- 1 & 0.4 -- 1 & 0 -- 10 & 1 -- 10 & $0$ -- $2\pi$ & $4\pi/5$ -- $6\pi/5$\\ \hline
$M_D^{(7)}$ & 0.3 -- 1 & 0.2 -- 1 & 0.5 -- 1 & 0.2 -- 1 & 0.3 -- 1 & 0 -- 10 & 1 -- 10 & $0$ -- $2\pi$ & $4\pi/5$ -- $6\pi/5$\\ \hline
\end{tabular}
\caption{Allowed parameter regions for IH cases of $M_D^{(2)}$ and $M_D^{(7)}$ in Eq.~(\ref{eq:Dirac}).}
\label{tab:IH}
\end{table}

\begin{figure}[h]
\begin{center}
\begin{tabular}{cc}
\hline 
\multicolumn{2}{c}{Texture $M_D^{(2)}$ in NH case} \\
\hline \\
\hspace{1.1cm}(a) & \hspace{1cm}(b) \\
\includegraphics[clip, width=0.45\hsize]{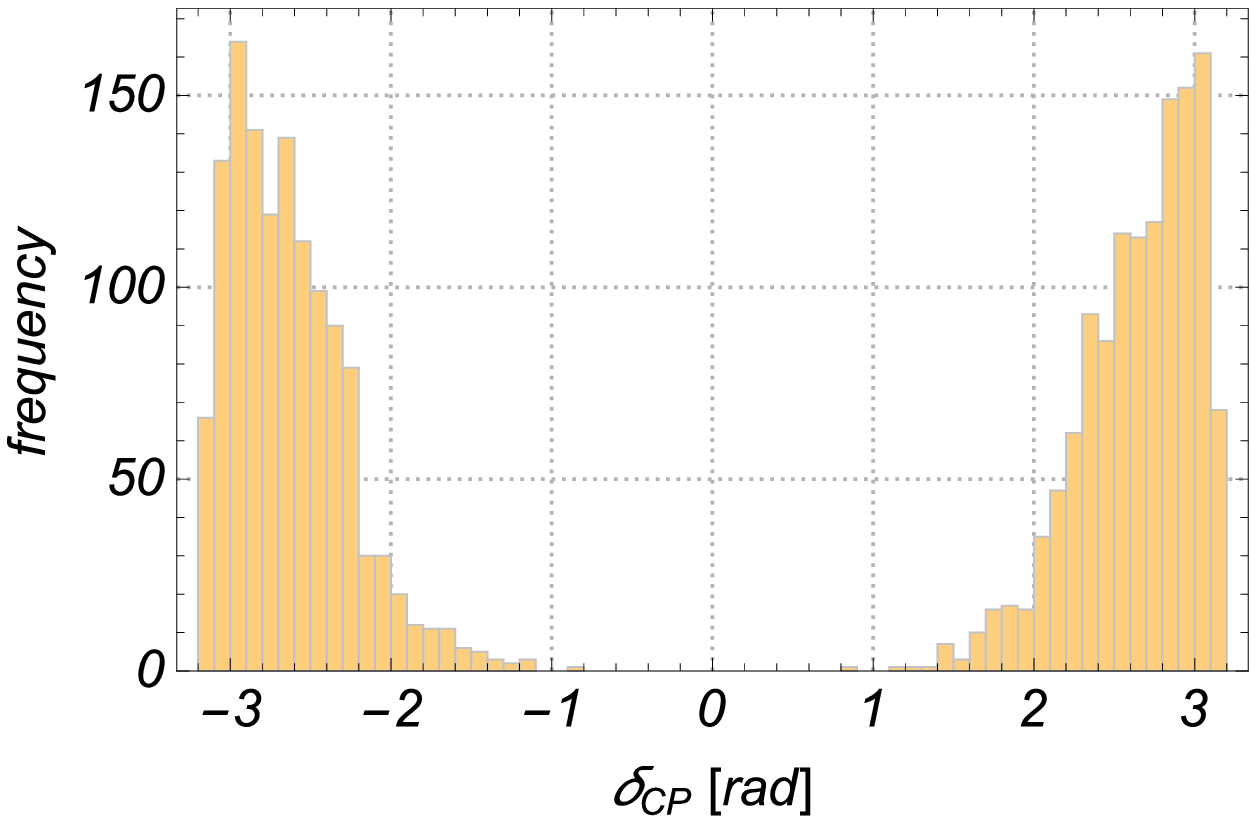} & \includegraphics[clip, width=0.45\hsize]{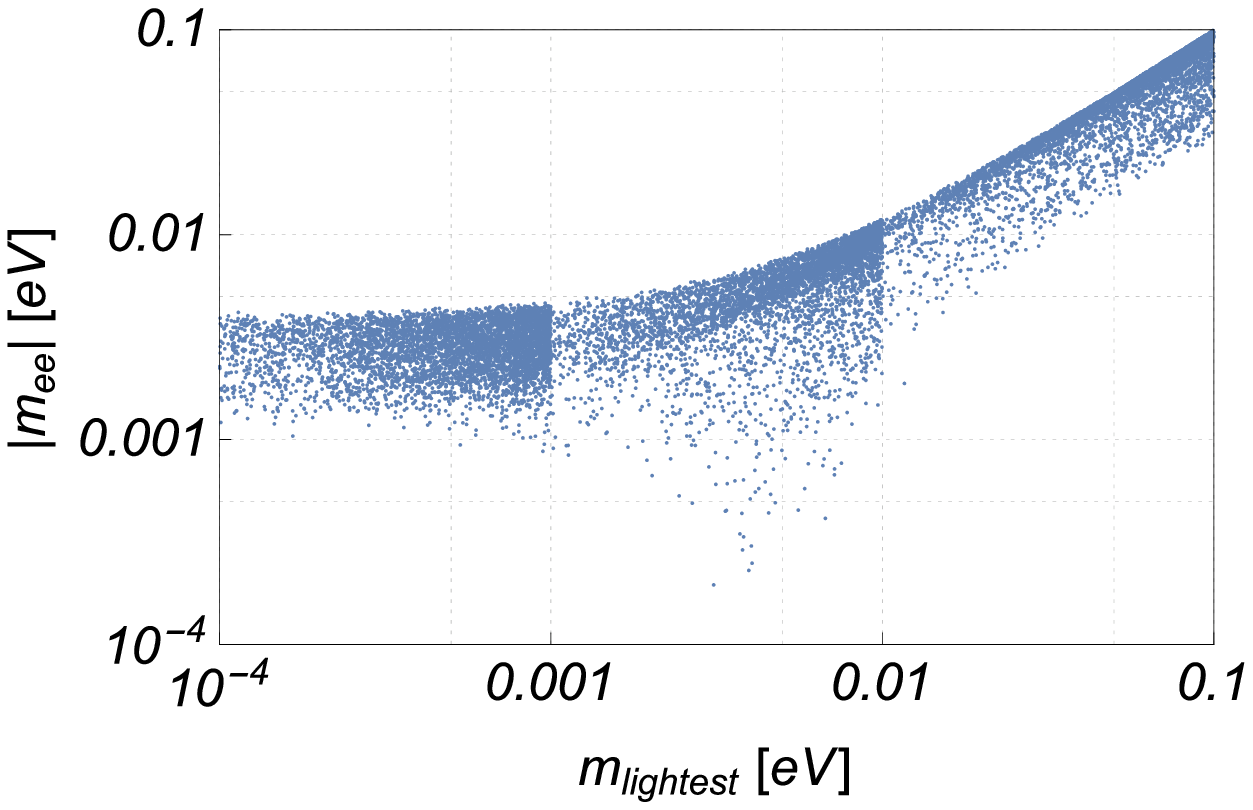} \\
\\
\hspace{1.1cm}(c) & \hspace{1cm}(d) \\
\includegraphics[clip, width=0.45\hsize]{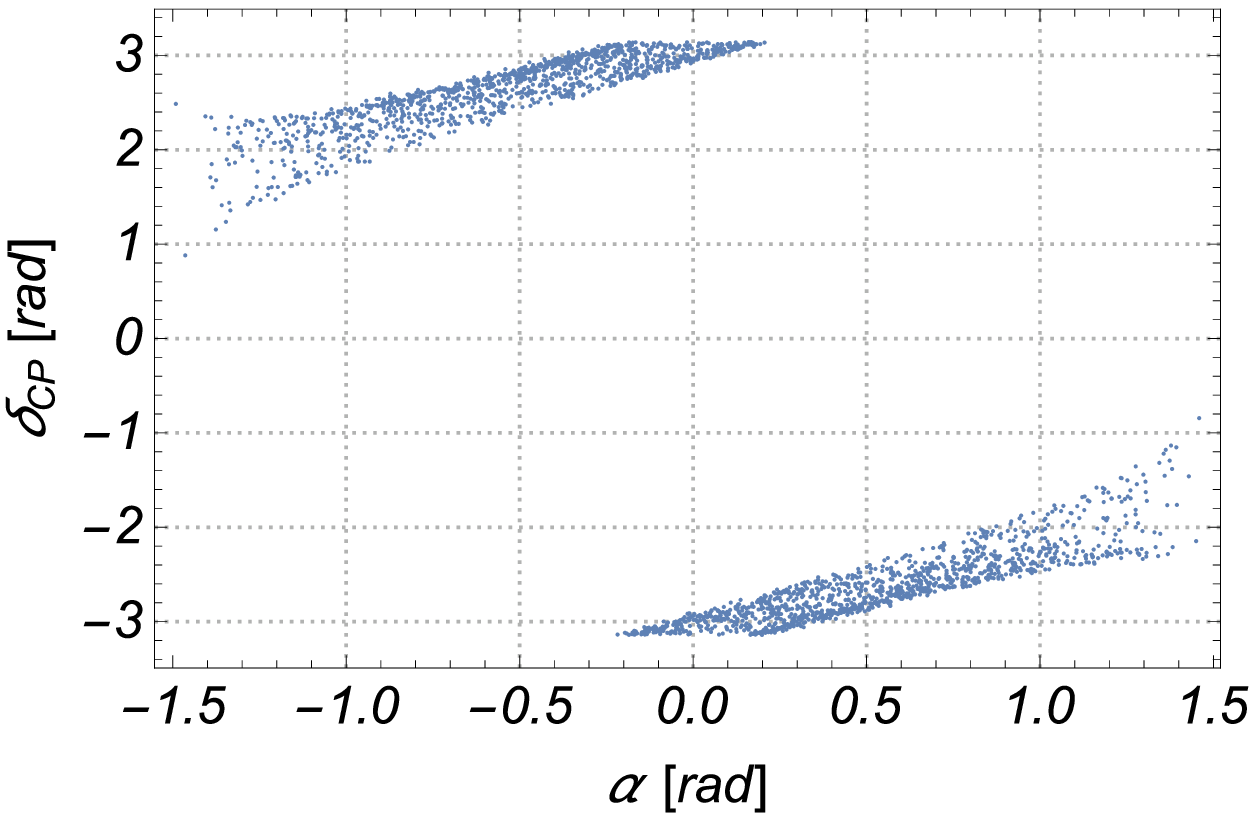} & \includegraphics[clip, width=0.45\hsize]{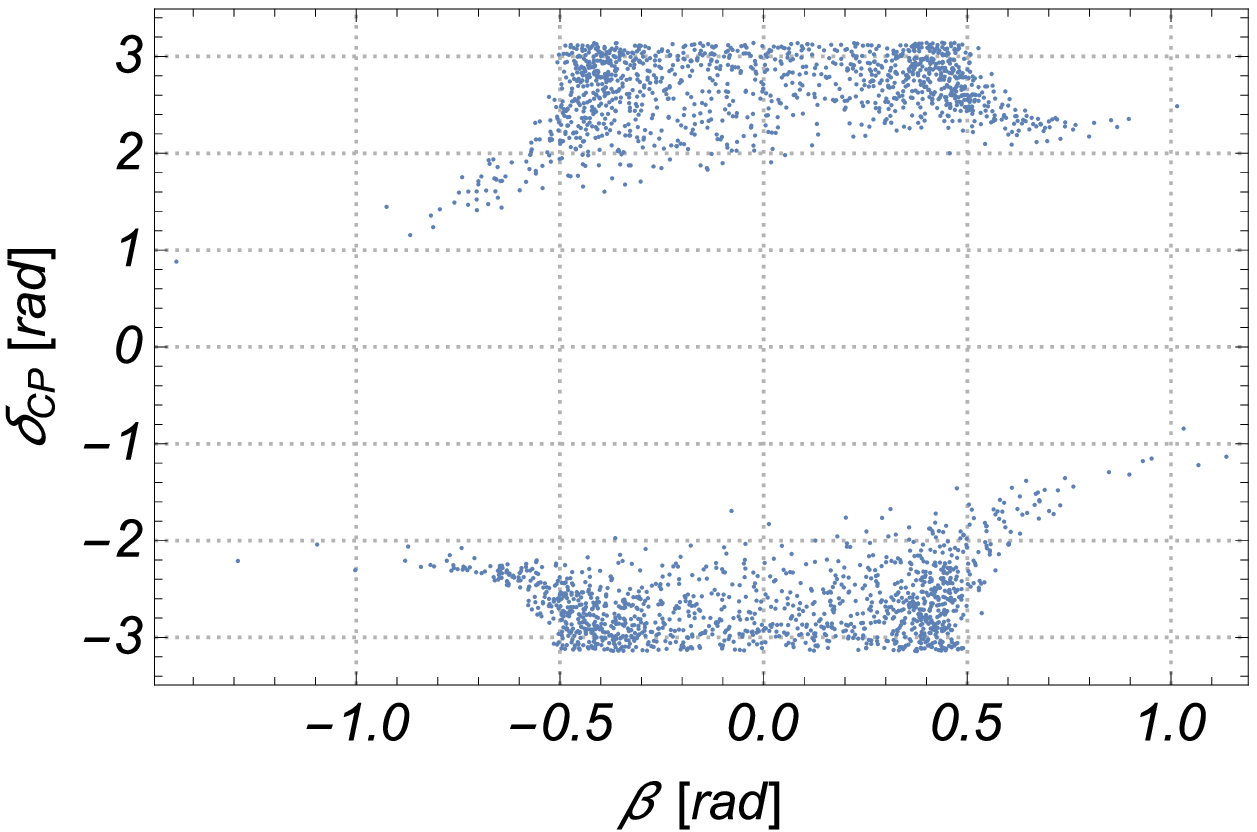} \\
\hline
\end{tabular}
\end{center}
\caption{The frequency distribution of the predicted Dirac CP violating phase $\delta _{CP}$, 
predicted regions of the effective mass for $0\nu \beta \beta $ decay, and Majorana phases $\alpha $ and $\beta $ 
in NH case of $M_D^{(2)}$: (a) frequency distribution of $\delta _{CP}$, (b) $m_{\text{lightest}}$--$|m_{ee}|$, (c) $\alpha $--$\delta _{CP}$, 
(d) $\beta $--$\delta _{CP}$. 
In the all figures from (a) to (d), the plots are shown within $3\sigma $ of $\sin ^2\theta _{12}$, $\sin ^2\theta _{23}$, $\sin ^2\theta _{13}$, and the ratio of the two neutrino mass squared differences in Eq.~\eqref{dataNH}.}
\label{fig:2NH}
\end{figure}

{\bf Texture $M_D^{(2)}$ in NH case.} We show the numerical results for NH case of $M_D^{(2)}$. The allowed parameter ranges for NH case are 
shown upward in Table~\ref{tab:NH}. In Fig.~\ref{fig:2NH}, we show the frequency distribution of the predicted Dirac CP violating phase $\delta _{CP}$, 
the predicted regions of the effective mass for $0\nu \beta \beta $ decay $|m_{ee}|$, and Majorana phases $\alpha $ and $\beta $. 
The frequency distribution of the $\delta _{CP}$ is shown in the Fig.~\ref{fig:2NH} (a). We found that $\delta _{CP}$ is predicted around $\pm 3$ rad. Then, it is testable 
for the T2K and NO$\nu $A experiments \cite{Abe:2013hdq, Adamson:2016tbq} in the near future. Indeed, T2K reported the first measurement of CP violating phase as $\delta _{CP}\simeq -\pi/2$ rad \cite{Abe:2013hdq}. 
If $\delta _{CP}$ is measured more precisely, the texture $M_D^{(2)}$ should be tested in NH case. 
In the Fig.~\ref{fig:2NH} (b), we show the effective mass for the $0\nu \beta \beta $ decay $|m_{ee}|$.\footnote{In numerical analyses for the effective mass, we use the observed values of neutrino mass squared differences \eqref{dataNH} without determining a value of the overall parameter $m_0$. The same holds for the following figures.} Since this texture $M_D^{(2)}$ is in NH case, $|m_{ee}|$ is too small 
to be measured except for the degenerate neutrino mass hierarchy case. 
In the Figs.~\ref{fig:2NH} (c) and (d), we show the predicted relations for the Dirac CP violating phase $\delta _{CP}$ versus Majorana phases $\alpha $ and $\beta $, respectively. 
The predicted relations imply that the Dirac CP violating phase $\pi/3~\text{rad}\lesssim \delta _{CP}\lesssim \pi ~\text{rad}$ corresponds to $-\pi /2 ~\text{rad}\lesssim \alpha \lesssim 0.2 ~\text{rad}$ and 
 $-\pi ~\text{rad}\lesssim \delta _{CP}\lesssim -\pi/3~\text{rad}$ corresponds to $-0.2~\text{rad}\lesssim \alpha \lesssim \pi /2~\text{rad}$, respectively. 
On the other hand, in the predicted ranges of the Dirac CP violating phase $-\pi ~\text{rad}\lesssim \delta _{CP}\lesssim -\pi/3~\text{rad}$ and $\pi/3~\text{rad}\lesssim \delta _{CP}\lesssim \pi ~\text{rad}$, 
the Majorana phase $\beta $ cannot be distinguished but we predict that the allowed range of $\beta $ is limited to $-\pi/3~\text{rad}\lesssim \beta \lesssim \pi/3~\text{rad}$. 

\begin{figure}[h]
\begin{center}
\begin{tabular}{cc}
\hline 
\multicolumn{2}{c}{Texture $M_D ^{(2)}$ in IH case} \\
\hline \\
\hspace{1.1cm}(a) & \hspace{1cm}(b) \\
\includegraphics[clip, width=0.45\hsize]{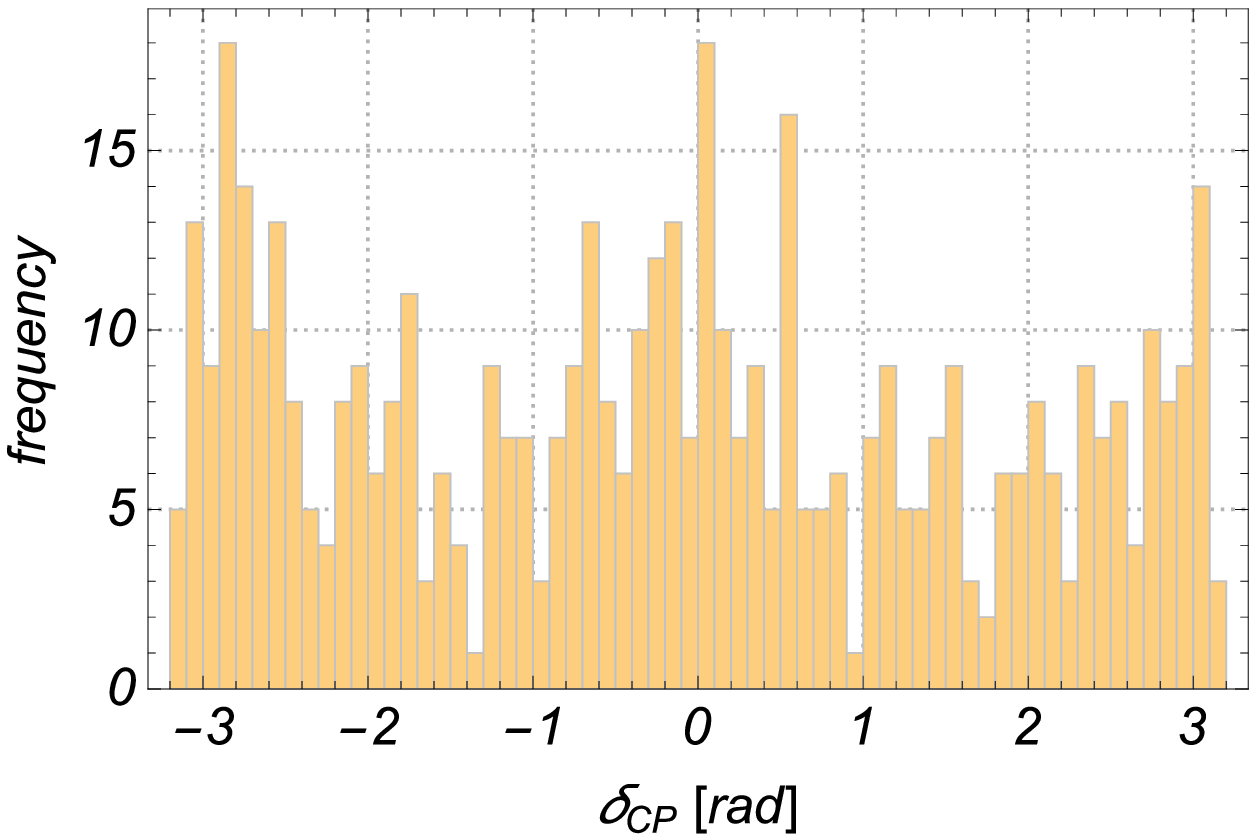} & \includegraphics[clip, width=0.45\hsize]{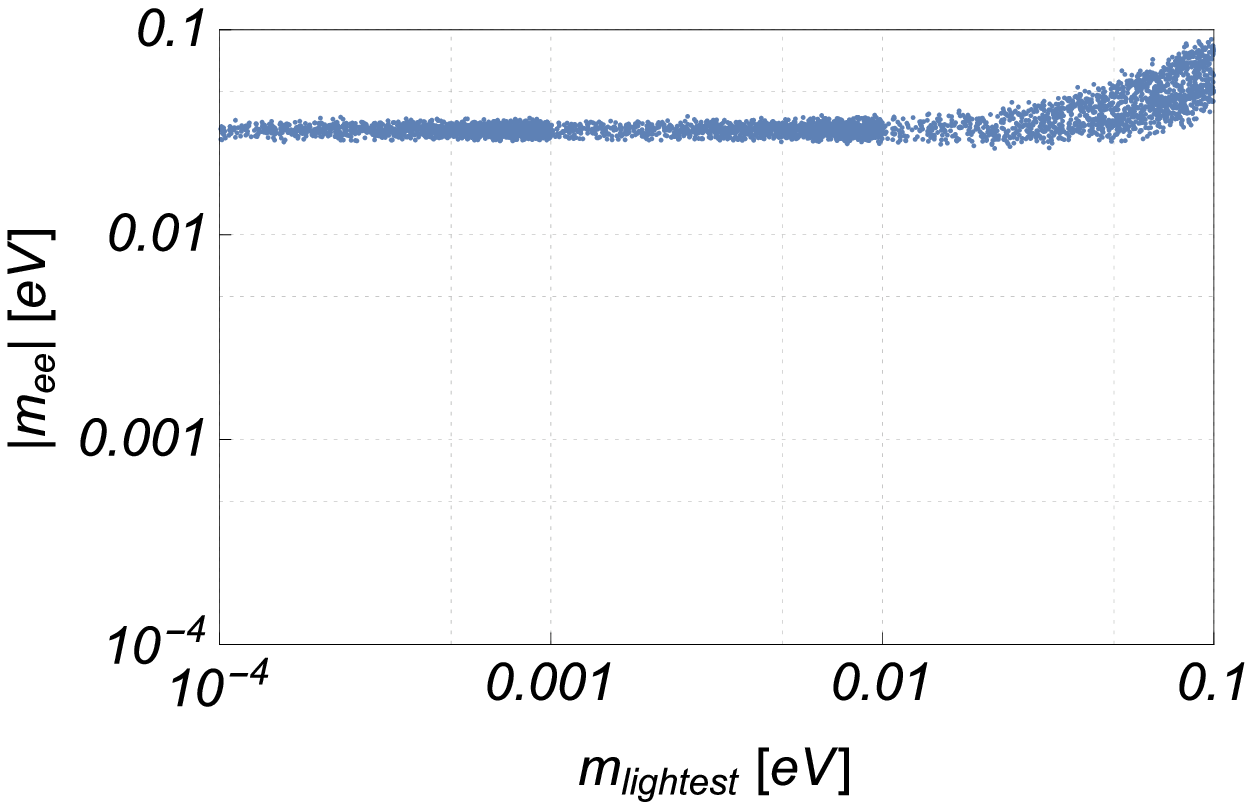} \\
\\
\hspace{1.1cm}(c) & \hspace{1cm}(d) \\
\includegraphics[clip, width=0.45\hsize]{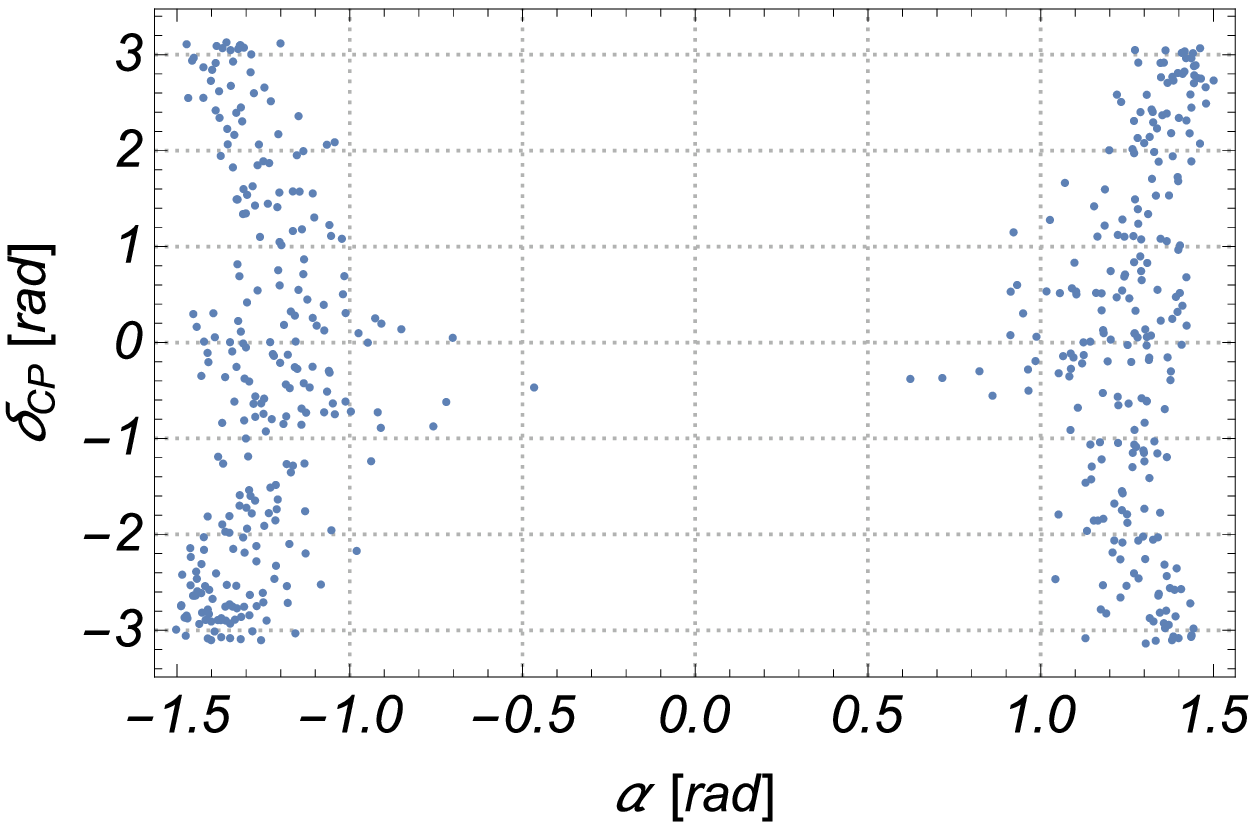} & \includegraphics[clip, width=0.45\hsize]{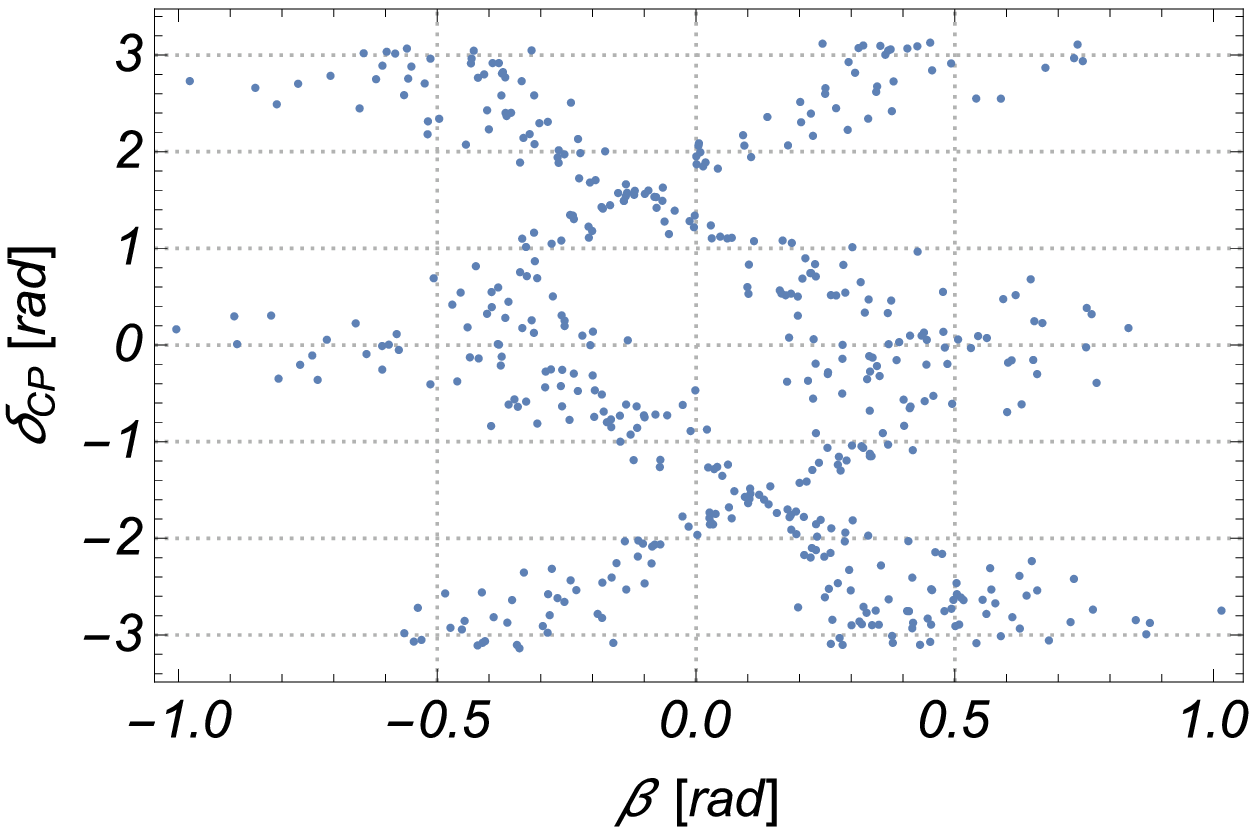} \\
\hline
\end{tabular}
\end{center}
\caption{The frequency distribution of the predicted Dirac CP violating phase $\delta _{CP}$, 
predicted regions of the effective mass for $0\nu \beta \beta $ decay, and Majorana phases $\alpha $ and $\beta $ 
in IH case of $M_D^{(2)}$: (a) frequency distribution of $\delta _{CP}$, (b) $m_{\text{lightest}}$--$|m_{ee}|$, (c) $\alpha $--$\delta _{CP}$, 
(d) $\beta $--$\delta _{CP}$. 
In the all figures from (a) to (d), the plots are shown within $3\sigma $ of $\sin ^2\theta _{12}$, $\sin ^2\theta _{23}$, $\sin ^2\theta _{13}$, and the ratio of the two neutrino mass squared differences in Eq.~\eqref{dataIH}.}
\label{fig:2IH}
\end{figure}

{\bf Texture $M_D^{(2)}$ in IH case.} We show the numerical results for NH case of $M_D^{(2)}$. The allowed parameter regions for IH case are 
shown upward in Table~\ref{tab:IH}. In Fig.~\ref{fig:2IH}, we show the frequency distribution of the predicted Dirac CP violating phase $\delta _{CP}$, 
the predicted regions of the effective mass for $0\nu \beta \beta $ decay $|m_{ee}|$, and Majorana phases $\alpha $ and $\beta $. 
The frequency distribution of the $\delta _{CP}$ is shown in the Fig.~\ref{fig:2IH} (a). We found that $\delta _{CP}$ is predicted in the all region. Then, it is difficult to test 
for the T2K and NO$\nu $A experiments in the near future. 
In the Fig.~\ref{fig:2IH} (b), we show the effective mass for the $0\nu \beta \beta $ decay $|m_{ee}|$. 
The predicted region of $|m_{ee}|$ is around $0.03 - 0.04$ eV, then it is testable for the KamLAND-Zen experiment \cite{KamLAND-Zen:2016pfg} in the near future. 
In the Figs.~\ref{fig:2IH} (c) and (d), we show the predicted regions for the Dirac CP violating phase $\delta _{CP}$ versus Majorana phases $\alpha $ and $\beta $, respectively. 
In IH case, the Majorana phase $\alpha $ is predicted as $0.5~\text{rad}\lesssim |\alpha |\lesssim \pi /2~\text{rad}$. 
On the other hand, the Majorana phase $\beta $ is predicted as $0~\text{rad}\lesssim |\beta |\lesssim \pi/3~\text{rad}$. 

\begin{figure}[h]
\begin{center}
\begin{tabular}{cc}
\hline 
\multicolumn{2}{c}{Texture $M_D^{(7)}$ in NH case} \\
\hline \\
\hspace{1.1cm}(a) & \hspace{1cm}(b) \\
\includegraphics[clip, width=0.45\hsize]{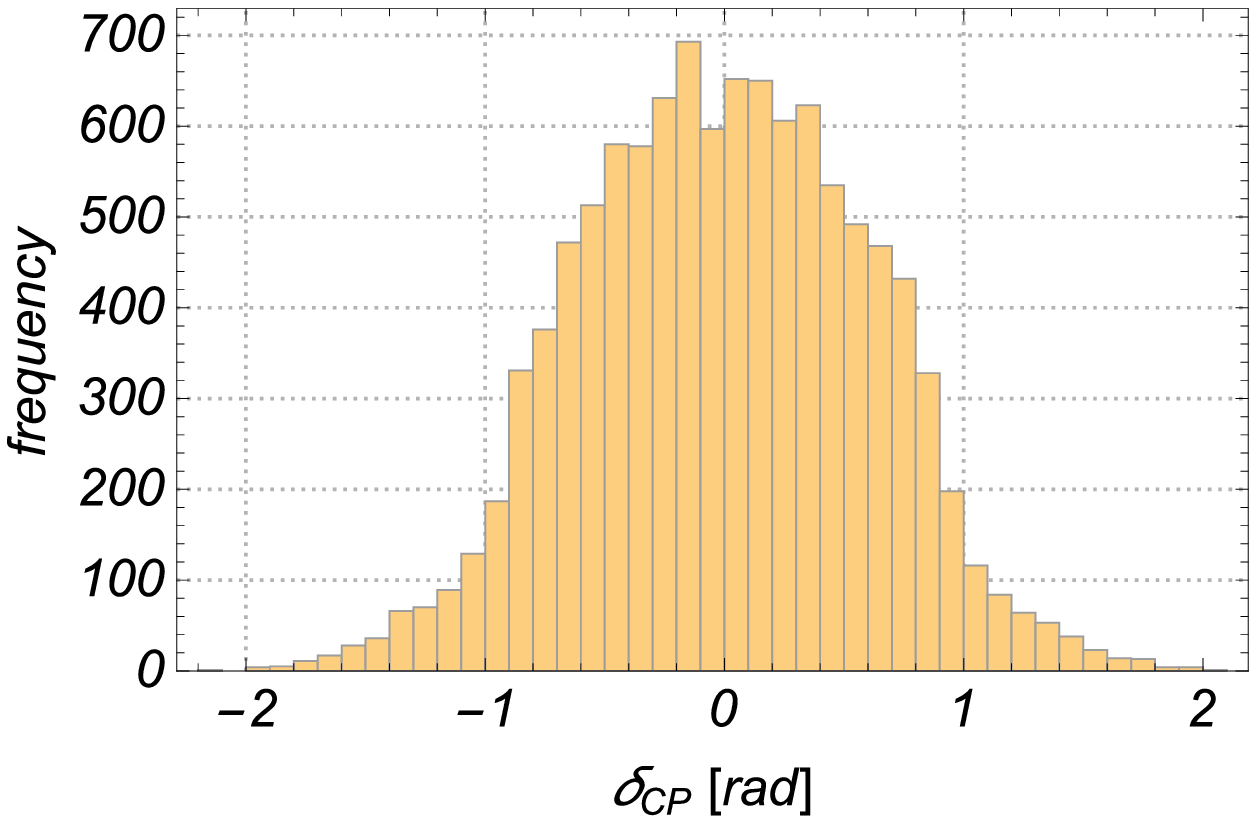} & \includegraphics[clip, width=0.45\hsize]{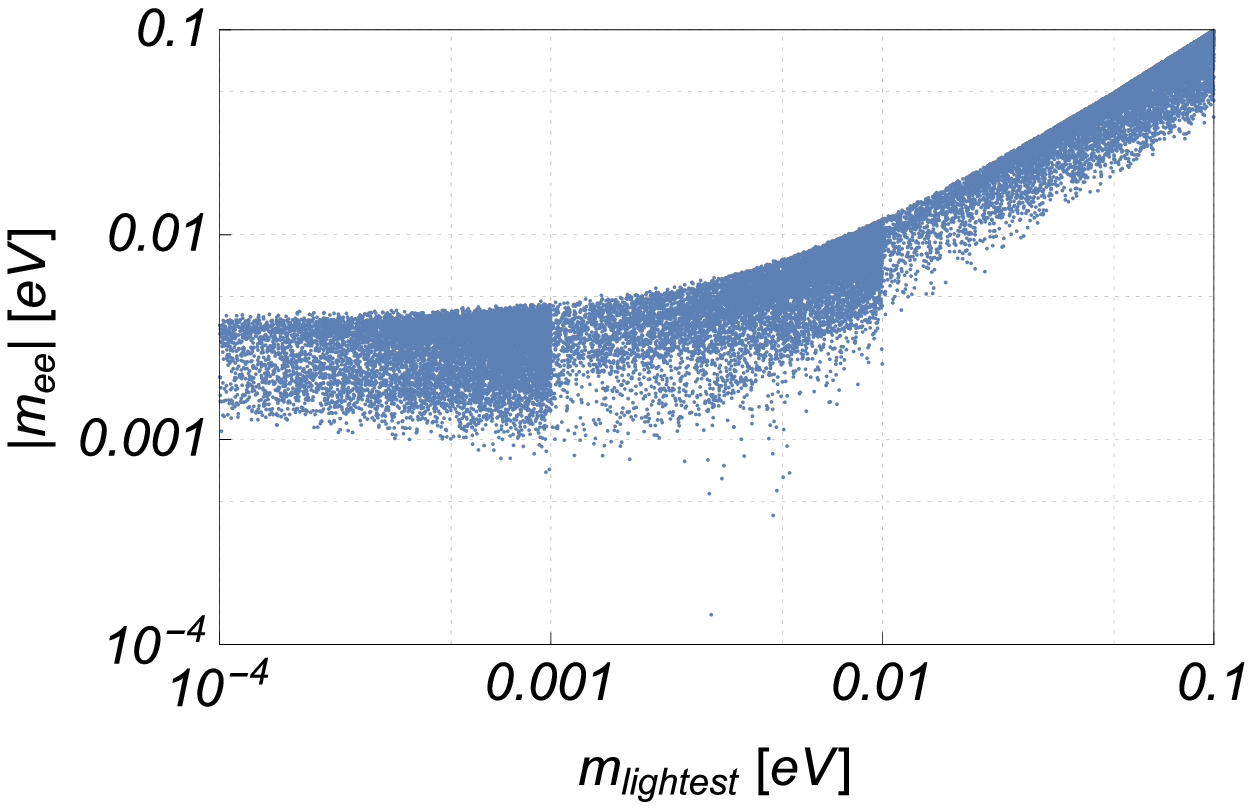} \\
\\
\hspace{1.1cm}(c) & \hspace{1cm}(d) \\
\includegraphics[clip, width=0.45\hsize]{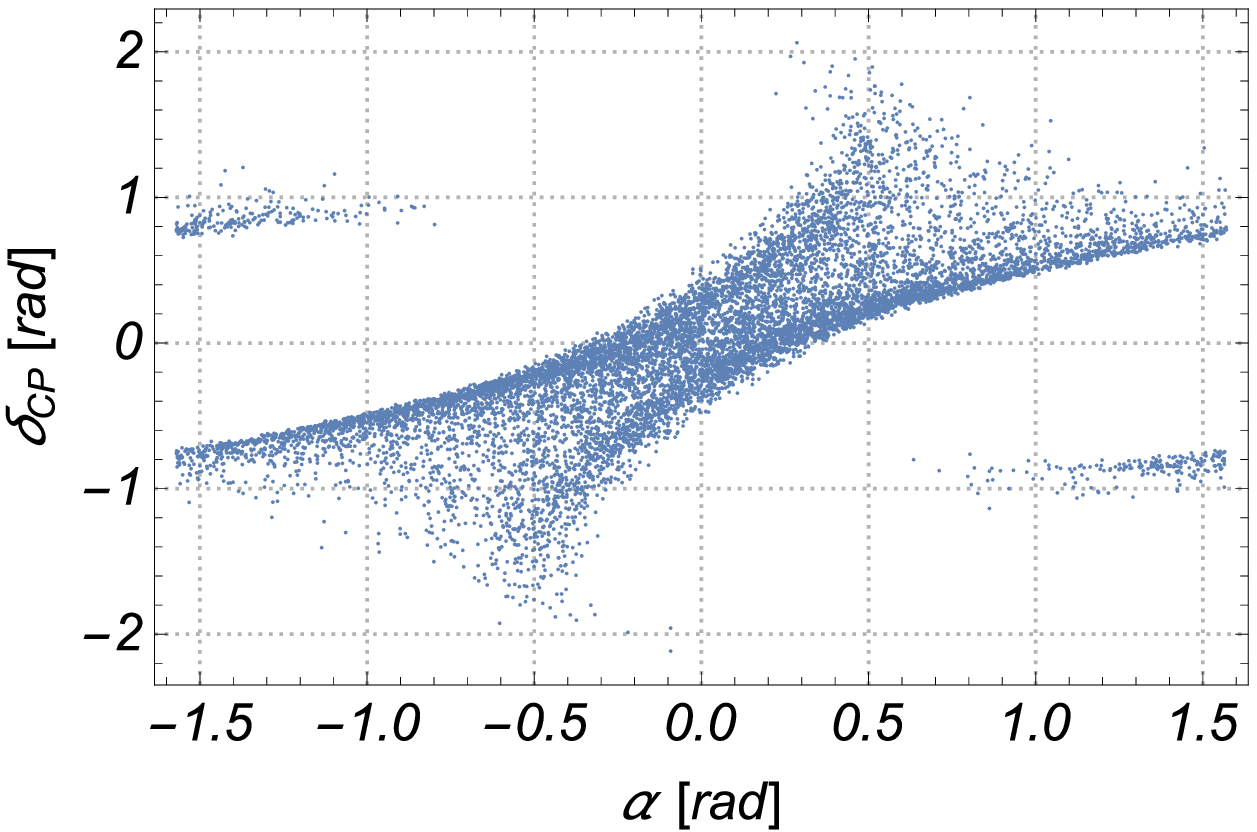} & \includegraphics[clip, width=0.45\hsize]{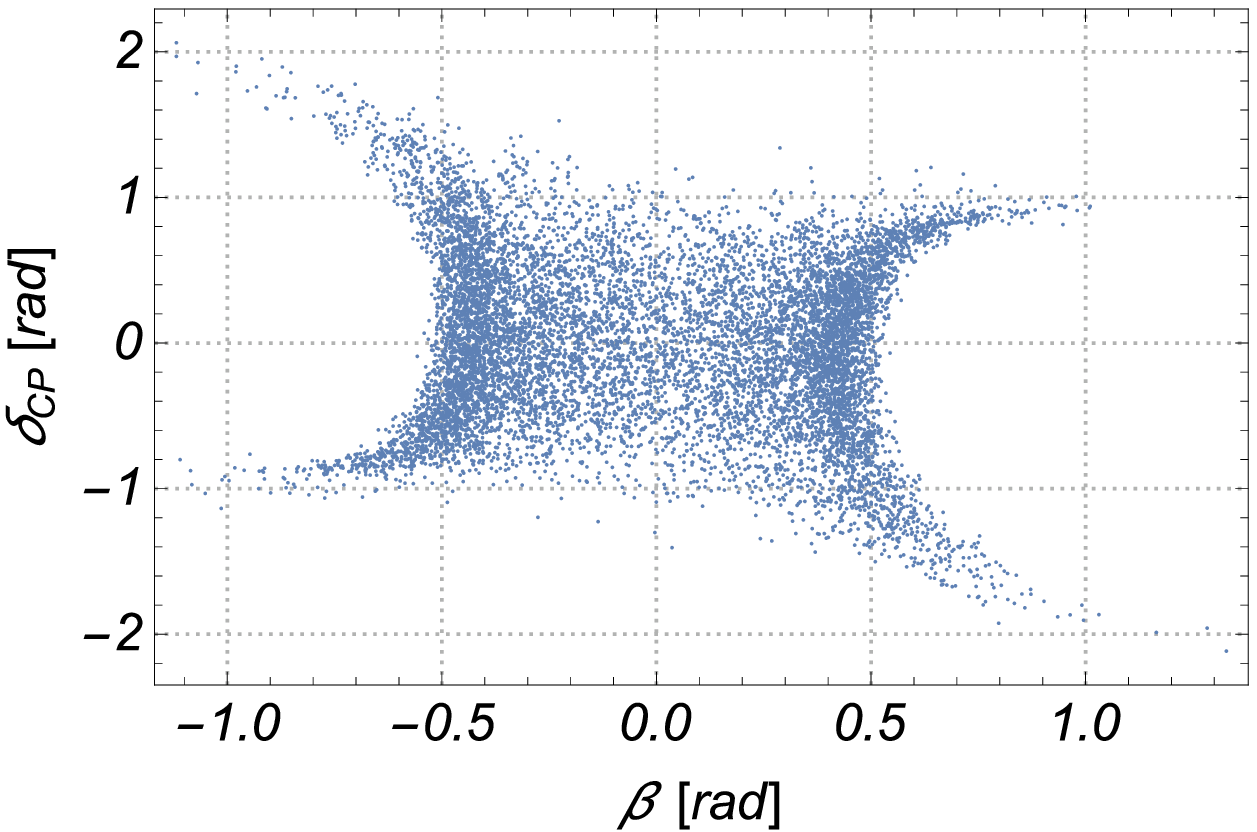} \\
\hline
\end{tabular}
\end{center}
\caption{The frequency distribution of the predicted Dirac CP violating phase $\delta _{CP}$, 
predicted regions of the effective mass for $0\nu \beta \beta $ decay, and Majorana phases $\alpha $ and $\beta $ 
in NH case of $M_D^{(7)}$: (a) frequency distribution of $\delta _{CP}$, (b) $m_{\text{lightest}}$--$|m_{ee}|$, (c) $\alpha $--$\delta _{CP}$, 
(d) $\beta $--$\delta _{CP}$. 
In the all figures from (a) to (d), the plots are shown within $3\sigma $ of $\sin ^2\theta _{12}$, $\sin ^2\theta _{23}$, $\sin ^2\theta _{13}$, and the ratio of the two neutrino mass squared differences in Eq.~\eqref{dataNH}.}
\label{fig:7NH}
\end{figure}

{\bf Texture $M_D^{(7)}$ in NH case.} We show the numerical results for NH case of $M_D^{(7)}$. The allowed parameter regions for NH case are 
shown downward in Table~\ref{tab:NH}. In Fig.~\ref{fig:7NH}, we show the frequency distribution of the predicted Dirac CP violating phase $\delta_{CP}$, 
the predicted regions of the effective mass for $0\nu \beta \beta $ decay $|m_{ee}|$, and Majorana phases $\alpha $ and $\beta $. 
The frequency distribution of the $\delta _{CP}$ is shown in the Fig.~\ref{fig:7NH} (a). We found that $\delta _{CP}$ is predicted as $0~\text{rad}\lesssim |\delta _{CP}|\lesssim 2\pi/3~\text{rad} $. 
Then, it is testable for the T2K and NO$\nu $A experiments in the near future. 
In the Fig.~\ref{fig:7NH} (b), we show the effective mass for the $0\nu \beta \beta $ decay $|m_{ee}|$. Since we consider NH case of $M_D^{(7)}$, $|m_{ee}|$ is too small 
to be measured except for the degenerate neutrino mass hierarchy case. This situation is similar to NH case of $M_D^{(2)}$. 
In the Figs.~\ref{fig:7NH} (c) and (d), we show the predicted relations for the Dirac CP violating phase $\delta _{CP}$ versus Majorana phases $\alpha $ and $\beta $, respectively. 
In the predicted range of the Dirac CP violating phase $0~\text{rad}\lesssim |\delta _{CP}|\lesssim 2\pi/3~\text{rad}$, the Majorana phases $\alpha $ and $\beta $ are predicted as 
$0~\text{rad}\lesssim |\alpha |\lesssim \pi /2~\text{rad}$ and $0~\text{rad}\lesssim |\beta |\lesssim \pi/3~\text{rad}$, respectively. 

\begin{figure}[h]
\begin{center}
\begin{tabular}{cc}
\hline 
\multicolumn{2}{c}{Texture $M_D ^{(7)}$ in IH case} \\
\hline \\
\hspace{1.1cm}(a) & \hspace{1cm}(b) \\
\includegraphics[clip, width=0.45\hsize]{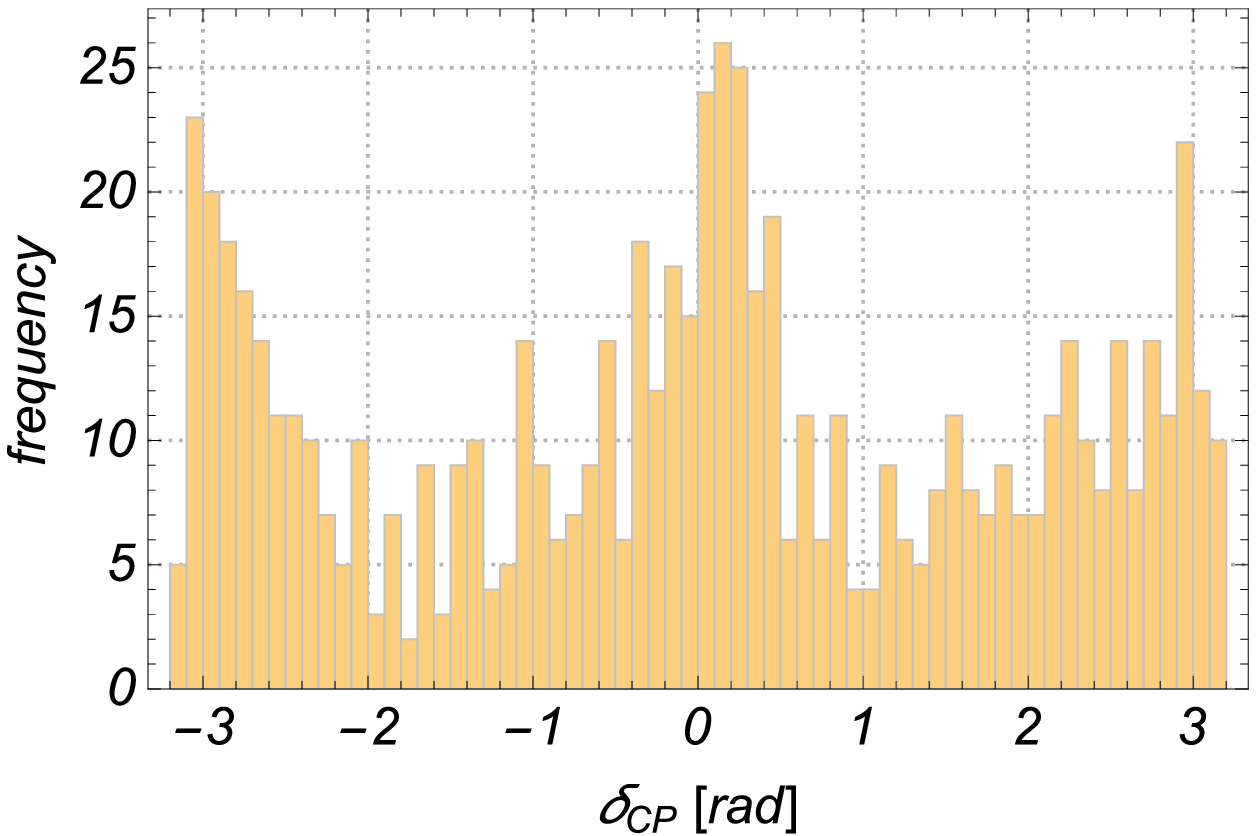} & \includegraphics[clip, width=0.45\hsize]{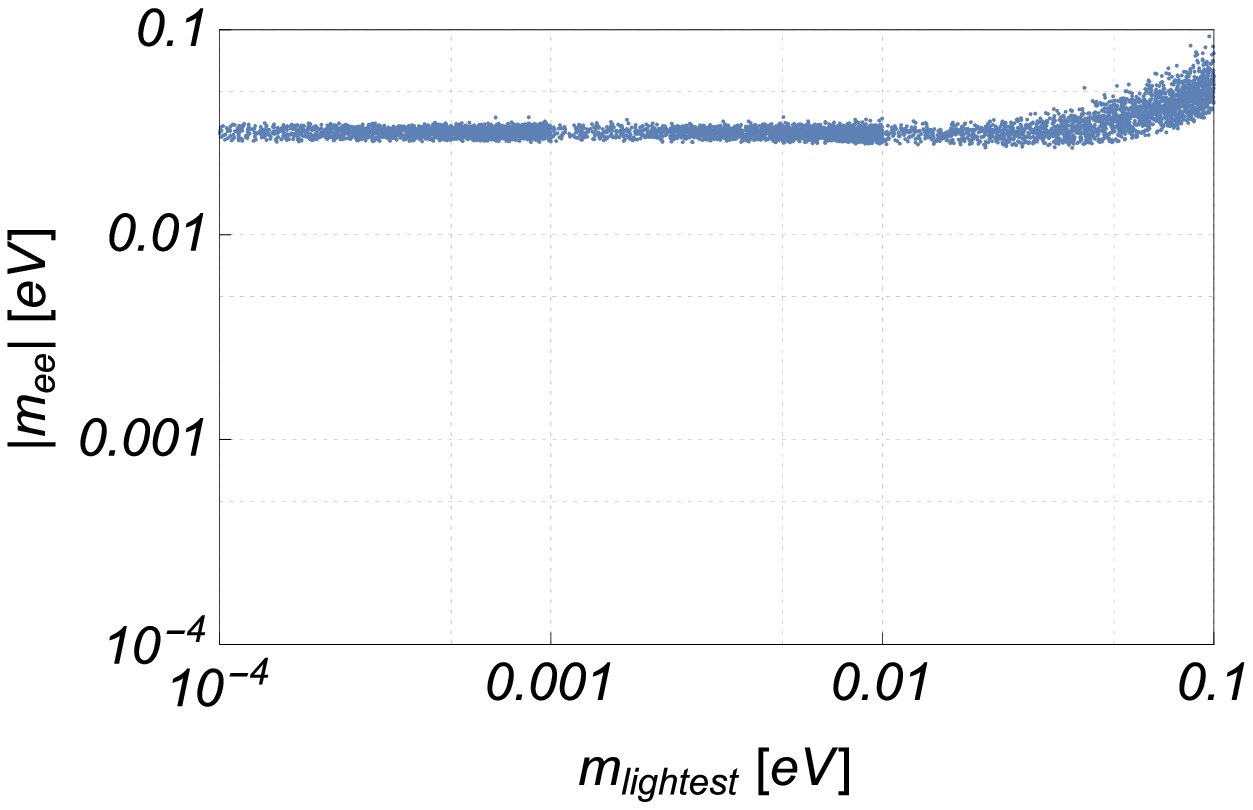} \\
\\
\hspace{1.1cm}(c) & \hspace{1cm}(d) \\
\includegraphics[clip, width=0.45\hsize]{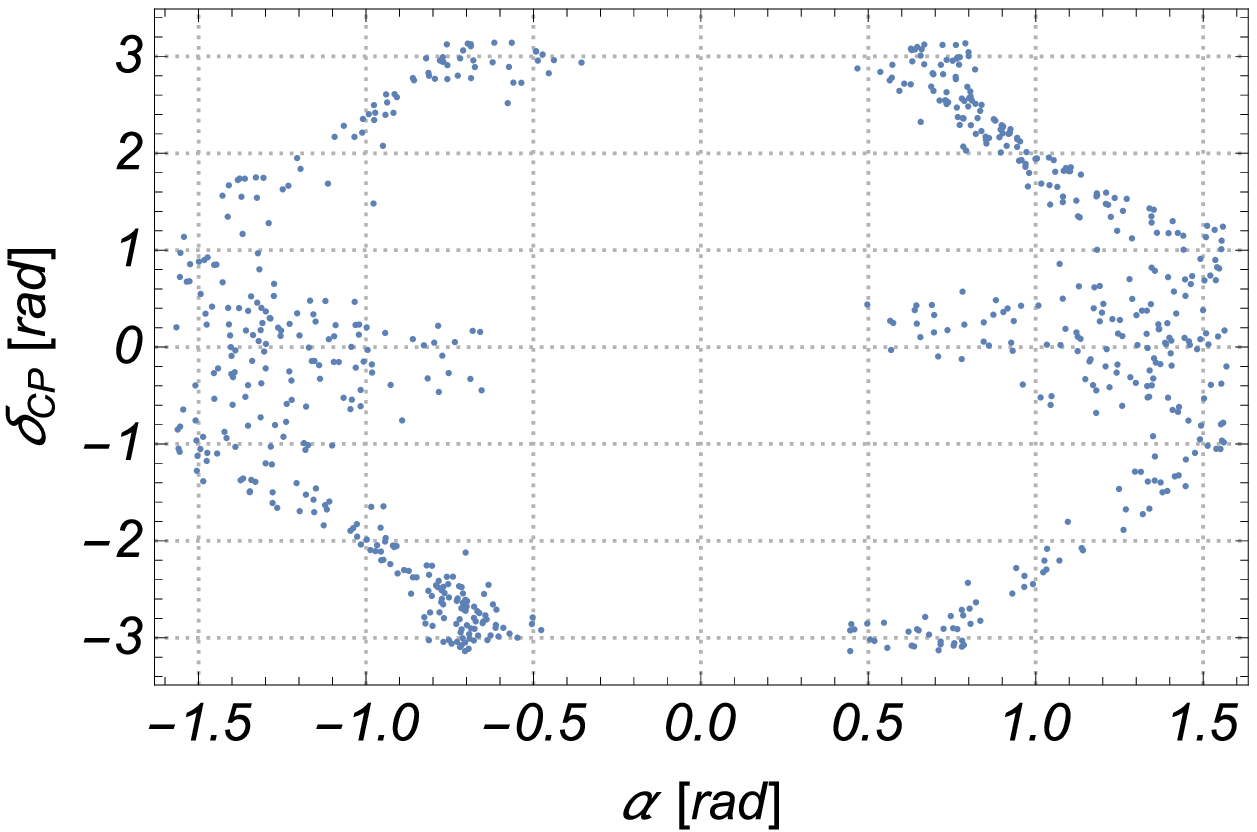} & \includegraphics[clip, width=0.45\hsize]{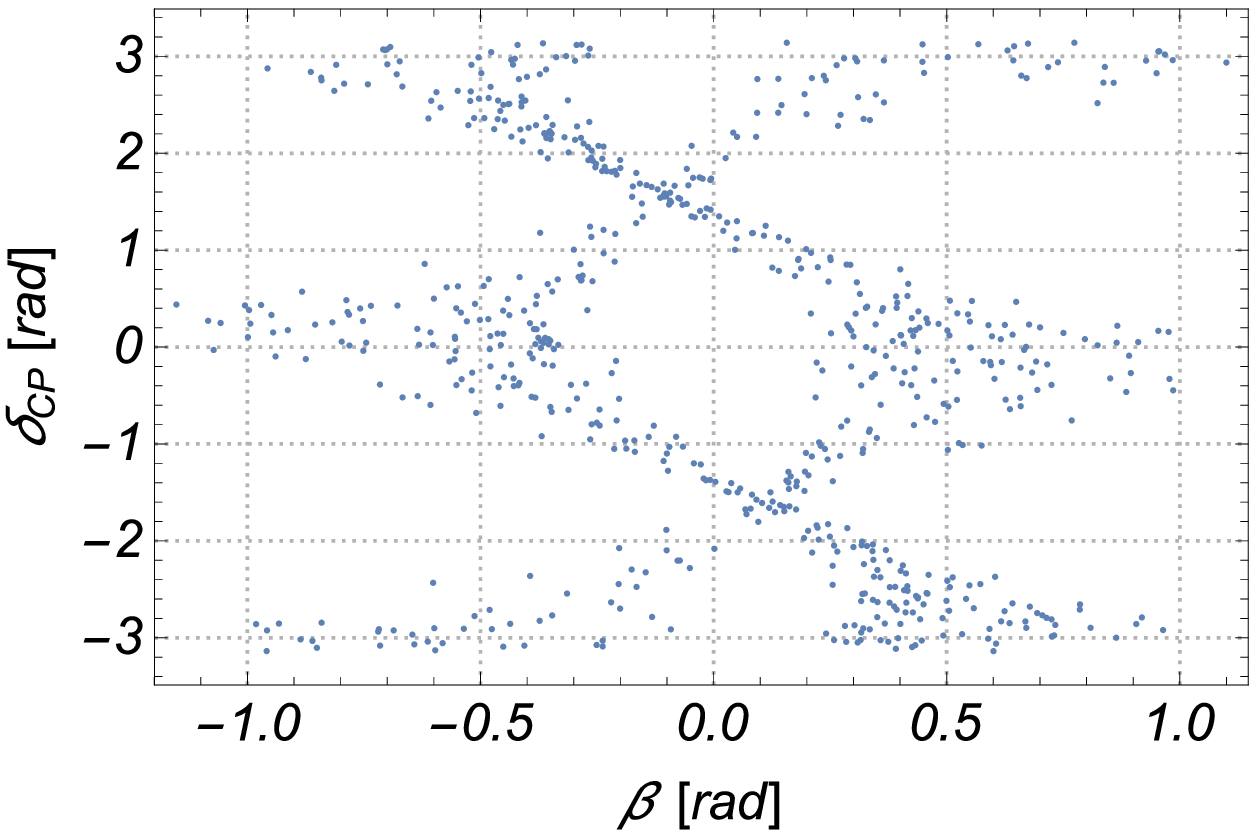} \\
\hline
\end{tabular}
\end{center}
\caption{The frequency distribution of the predicted Dirac CP violating phase $\delta _{CP}$, 
predicted regions of the effective mass for $0\nu \beta \beta $ decay, and Majorana phases $\alpha $ and $\beta $ 
in IH case of $M_D^{(7)}$: (a) frequency distribution of $\delta _{CP}$, (b) $m_{\text{lightest}}$--$|m_{ee}|$, (c) $\alpha $--$\delta _{CP}$, 
(d) $\beta $--$\delta _{CP}$. 
In the all figures from (a) to (d), the plots are shown within $3\sigma $ of $\sin ^2\theta _{12}$, $\sin ^2\theta _{23}$, $\sin ^2\theta _{13}$, and the ratio of the two neutrino mass squared differences in Eq.~\eqref{dataIH}.}
\label{fig:7IH}
\end{figure}

{\bf Texture $M_D^{(7)}$ in IH case.} We show the numerical results for IH case of $M_D^{(7)}$. The allowed parameter regions for IH case are 
shown downward in Table~\ref{tab:IH}. In Fig.~\ref{fig:7IH}, we show the frequency distribution of the predicted Dirac CP violating phase $\delta _{CP}$, 
the predicted regions of the effective mass for $0\nu \beta \beta $ decay $|m_{ee}|$, and Majorana phases $\alpha $ and $\beta $. 
The frequency distribution of the $\delta _{CP}$ is shown in the Fig.~\ref{fig:7IH} (a). We found that $\delta _{CP}$ is predicted in the all region. Then, it is difficult to test 
for the T2K and NO$\nu $A experiments in the near future. This situation is similar to IH case of $M_D^{(2)}$. 
In the Fig.~\ref{fig:7IH} (b), we show the effective mass for the $0\nu \beta \beta $ decay $|m_{ee}|$. 
The predicted region of $|m_{ee}|$ is around $0.03-0.04$ eV, then it is testable for the KamLAND-Zen experiment in the near future. 
In the Figs.~\ref{fig:7IH} (c) and (d), we show the predicted relations for the Dirac CP violating phase $\delta _{CP}$ versus Majorana phases $\alpha $ and $\beta $, respectively. 
The allowed ranges of Majorana phases $\alpha $ and $\beta $ are shown as $0.5~\text{rad}\lesssim |\alpha |\lesssim \pi /2~\text{rad}$ and 
$0~\text{rad}\lesssim |\alpha |\lesssim \pi/3~\text{rad}$, respectively.

\section{Discussions and Summary}
\label{sec:summary}
In this paper, we have discussed the neutrino flavor structures in the Occam's razor approach for the Dirac neutrino mass matrices. 
The Occam's razor approach is that we used a minimal set of non-zero entries in the Dirac mass matrices.
We assumed that the right-handed Majorana neutrino mass matrix is diagonal and we considered nine patterns of the four zero textures for the Dirac neutrino mass matrices. 
We looked at experimental and observational testabilities of the Occam's razor approach from the viewpoints of the CP violation and the effective mass in neutrinoless double beta decay.
For the two of the nine patterns, we numerically analyzed phenomenological aspects for the Dirac neutrino mass matrices, and as a result we found that our setups can lead to both of normal and inverted hierarchies of neutrino masses, depending on configurations of input parameter ranges. 
In a case of normal hierarchy, we found characteristic predictions in the CP violating phase, and it was also proved that it is likely that our setups will be verified by neutrino experiments, e.g., T2K and NO$\nu$A experiments, in the near future. On the other hand, in a case of inverted hierarchy, the effective mass in neutrinoless double beta decay was predicted to be relatively large compared with that in the case of normal hierarchy. It will be expected that our setups also in the case of inverted hierarchy can be verified by neutrino measurements, e.g., KamLAND-Zen, within a few years.
From the left-handed effective neutrino mass matrix, we determined concrete values of input mass parameters which can lead to the observed neutrino mass squared differences and mixing angles.
As stated in the previous section, in particular the allowed range of a phase $\phi_B$ in the right-handed Majorana mass matrix are strongly restricted.
Thus, it is interesting to apply our setups to models for the baryon asymmetry of the universe from the right-handed Majorana masses, i.e., leptogenesis.
That is left for the future work.

\vspace{0.3cm}
\noindent
{\bf Acknowledgement}

This work is supported in part by Grants-in-Aid for Scientific Research (No.~16J05332 (YS) and No.~16J04612 (YT)) from the Ministry of Education, Culture, Sports, Science and Technology in Japan.

\end{document}